\documentclass[prl,aps,amsfonts,amsmath,amssymb,nofootinbib,twocolumn,superscriptaddress]{revtex4-2}
\usepackage{graphicx}
\usepackage{bm}
\usepackage{hyperref}
\hypersetup{
    colorlinks,
    linkcolor={red!50!black},
    citecolor={blue!50!black},
    urlcolor={blue!80!black}
}
\usepackage{float}
\usepackage{xcolor}
\graphicspath{{figures/}}
\begin{document}
\title{Energy relaxation and dynamics in the correlated metal Sr$_2$RuO$_4$ via THz two-dimensional coherent spectroscopy}

\author{David Barbalas}
\affiliation{William H. Miller III, Department of Physics and Astronomy, The Johns Hopkins University, Baltimore, Maryland 21218, USA}
  
\author{Ralph Romero III}
\affiliation{William H. Miller III, Department of Physics and Astronomy, The Johns Hopkins University, Baltimore, Maryland 21218, USA}
  
\author{Dipanjan Chaudhuri}
\affiliation{William H. Miller III, Department of Physics and Astronomy, The Johns Hopkins University, Baltimore, Maryland 21218, USA}
  
\author{Fahad Mahmood}
\affiliation{William H. Miller III, Department of Physics and Astronomy, The Johns Hopkins University, Baltimore, Maryland 21218, USA}
\affiliation{Department of Physics, University of Illinois at Urbana-Champaign, Urbana, 61801 IL, USA}
\affiliation{F. Seitz Materials Research Laboratory, University of Illinois at Urbana-Champaign, Urbana, 61801 IL, USA}

\author{Hari P. Nair}
\affiliation{Department of Materials Science and Engineering, Cornell University, Ithaca, NY 14853, USA}

\author{Nathaniel J. Schreiber}
\affiliation{Department of Materials Science and Engineering, Cornell University, Ithaca, NY 14853, USA}

\author{Darrell G. Schlom}
\affiliation{Department of Materials Science and Engineering, Cornell University, Ithaca, NY 14853, USA}
\affiliation{Kavli Institute at Cornell for Nanoscale Science, Ithaca, New York 14853, USA}
\affiliation{Leibniz-Institut für Kristallzüchtung, Max-Born-Str. 2, 12489 Berlin, Germany}
 
\author{K. M. Shen}
\affiliation{Laboratory of Atomic and Solid State Physics, Department of Physics,
	Cornell University, Ithaca, New York 14853, USA} 
 
\author{N. P. Armitage}\email{npa@jhu.edu}
\affiliation{William H. Miller III, Department of Physics and Astronomy, The Johns Hopkins University, Baltimore, Maryland 21218, USA}
\affiliation{Canadian Institute for Advanced Research, Toronto, Ontario M5G 1Z8, Canada}

\date{\today}
\begin{abstract}
\textbf{Separating out the contributions of different scattering channels in strongly interacting metals is crucial in identifying the mechanisms that govern their properties. While momentum or current relaxation rates can be readily probed via \textit{dc} resistivity or optical/THz spectroscopy, distinguishing different kinds of inelastic scattering can be more challenging. Using nonlinear THz 2D coherent spectroscopy, we measure the rates of energy relaxation after THz excitation in the strongly interacting Fermi liquid, Sr$_2$RuO$_4$.  Energy relaxation is a bound on the total scattering and specifically a measure of contributions to the electron self-energy that arise from {\it inelastic} coupling to a bath.  We observe two distinct energy relaxation channels: a fast process that we interpret as energy loss to the phonon system and a much slower relaxation that we interpret as arising from a non-equilibrium phonon effects and subsequent heat loss through diffusion. Interestingly, even the faster energy relaxation rate is at least an order of magnitude slower than the overall momentum relaxation rate, consistent with strong electron interactions and the dominance of energy-conserving umklapp or interband electron-electron scattering in momentum relaxation. The slowest energy relaxation rate decays on a sub-GHz scale, consistent with the relaxation dynamics of non-equilibrium phonons. Our observations reveal the versatility of nonlinear THz spectroscopy to measure the energy relaxation dynamics in correlated metals.   Our work also highlights the need for improved theoretical understanding of such processes in interacting metals. }
\end{abstract}

\maketitle

Physical systems may be characterized by different rates of relaxation for various perturbations. The interrelation between the different rates may be interesting as they can give unique insight into microscopic processes. While momentum relaxation in metals can be readily probed via transport and optical experiments, isolating other relaxation channels can be challenging. Momentum relaxation is dominated by large angle momentum-relaxing scattering, but transport in metals may also be sensitive to other rates that govern dephasing, energy transport, energy loss, and thermalization. For instance, thermal and thermoelectric transport compliments charge transport in providing information that is more sensitive to all scattering  and the role of phonons due to their sensitivity to different scattering mechanisms~\cite{cutler_observation_1969,hwang_theory_2009}. One possibility to determine the relative magnitude of elastic and inelastic scattering in weakly interacting metals is the Wiedemann-Franz law (WFL), which relates electrical and heat transport.  Unfortunately, experimental challenges isolating the electronic contribution to heat transport can limit the applicability of this technique to a narrow range of temperatures~\cite{grissonnanche_wiedemann-franz_2016,grissonnanche_giant_2019}. The thermoelectric response can also host signatures of strong electron-phonon coupling, such as phonon drag observed in clean metals at low temperatures. In that case, strong electron-phonon coupling transfers momentum into the phonon system faster than it can relax out, driving the phonon distribution out of equilibrium. This modifies both the thermal and electrical transport properties from what is expected in equilibrium~\cite{herring_theory_1954,r_n_gurzhi_electric_1972,yu_g_gurevich_electron-phonon_1989}. While the effect on resistivity is typically small, phonon drag has a large impact on the thermopower~\cite{behnia_thermoelectricity_2004,quan_impact_2021}.

\begin{figure*}[!ht]
    \centering
    \includegraphics[width = 1.0 \textwidth]{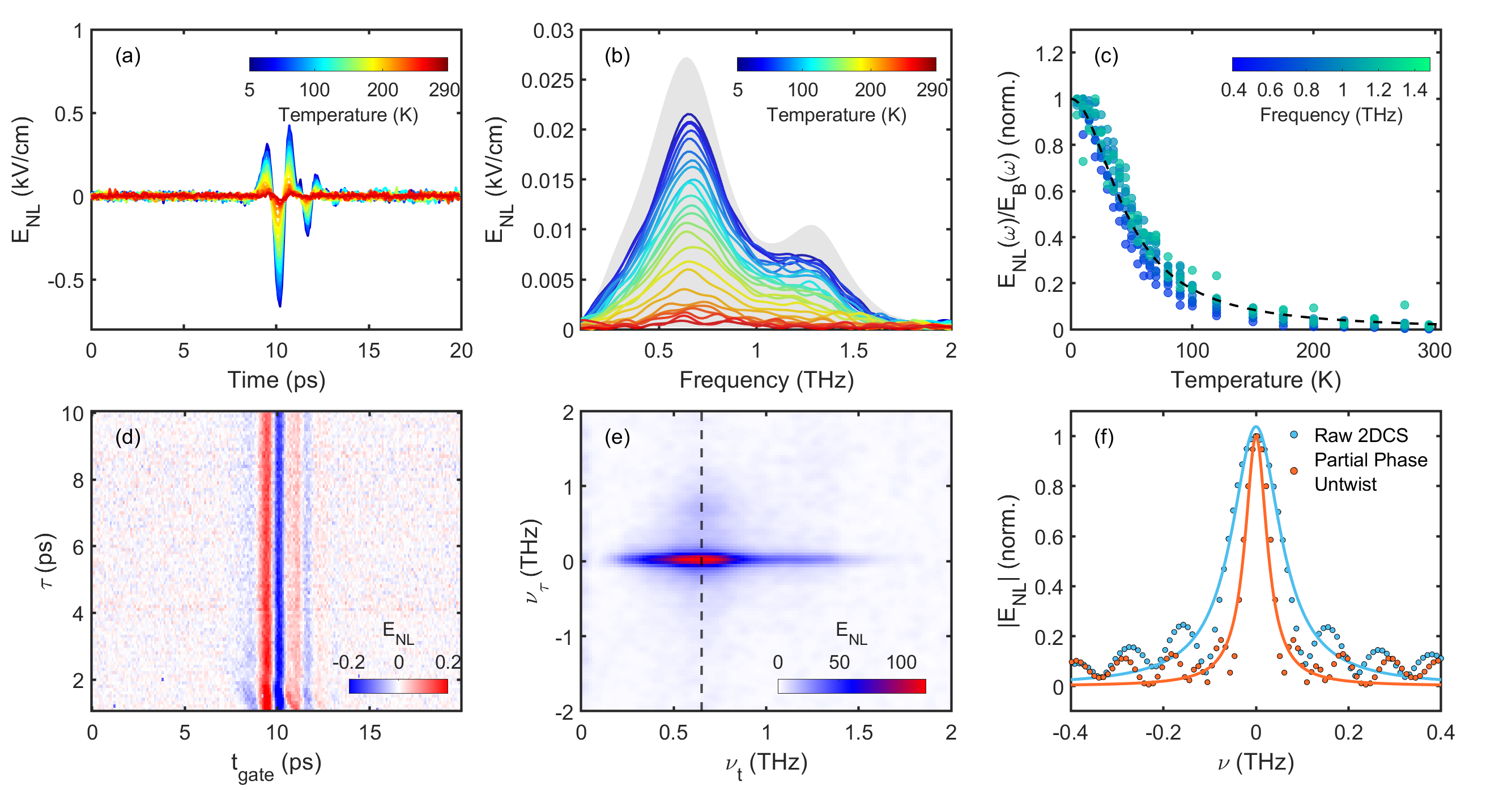}
    \caption{(a) The temperature dependence of the THz nonlinear response of Sr$_2$RuO$_4$ and (b) the corresponding Fourier transform. The applied electric field through the aperture is shown as a gray outline for reference (scaled by 0.05). The normalized temperature dependence $E_{NL}/E_B$ along different frequency cuts is shown in (c), with a Lorentzian guide to the eye with a half-width at half-maximum of $44$ K $\sim 1$ THz.     (d,e)  THz 2DCS response at 10 K in the cross-polarized regime in the time- and frequency-domains. (f) Cut of the raw 2DCS Fourier transform and partially phase untwisted spectra at the position marked by dashed line along with corresponding fits giving $\Gamma = 0.064(6)$ THz and  $\Gamma = 0.029(6)$ THz.}
    \label{Fig1_2pulse}
\end{figure*}

The electronic energy relaxation rate characterizes the rate that energy leaves the electron system after excitation. Energy relaxes ultimately into the subsystem with the largest heat capacity. In normal metals this is the lattice, and so energy relaxation is dominated by inelastic electron-phonon scattering. A common model for understanding energy relaxation is the two-temperature model (TTM), where one assumes that the energy injected into the electronic system quickly relaxes to a quasi-thermal distribution characterized by a temperature $T_e$ and then more slowly loses energy to a bath that is at a temperature $T_L$~\cite{allen_theory_1987,groeneveld1995femtosecond}. Due to a relatively universal cancellation at low $T$ that arises from the fact that both the electronic system's heat capacity and the typical emitted phonon energy goes at $T$, one can show that the energy relaxation rate is a generic {\it upper} bound on the total quasiparticle scattering rate. For the case of energy loss via acoustic phonons, Allen found~\cite{allen_theory_1987} explicitly that the energy relaxation $\Gamma_E  = 12  \zeta(3)  \lambda \Big [ \frac{k_B T_e}{\hbar \omega_D} \Big ]^3  \omega_D$, where $\zeta$ is the Riemann zeta function and $\omega_D$ is the Debye frequency.  One can compare this to the phononic contribution to particle-hole relaxation rate $\Gamma_\sigma$ that calculations~\cite{allen1975comment} have shown to be related to energy relaxation as $\Gamma_E / \Gamma_\sigma \approx 1.31 $, e.g. energy relaxation is a bound on the total scattering and specifically a measure of contributions to the self-energy that arise from {\it inelastic} coupling to the bath.

\begin{figure*}[!ht]
    \centering
    \includegraphics[width = 1.0 \textwidth]{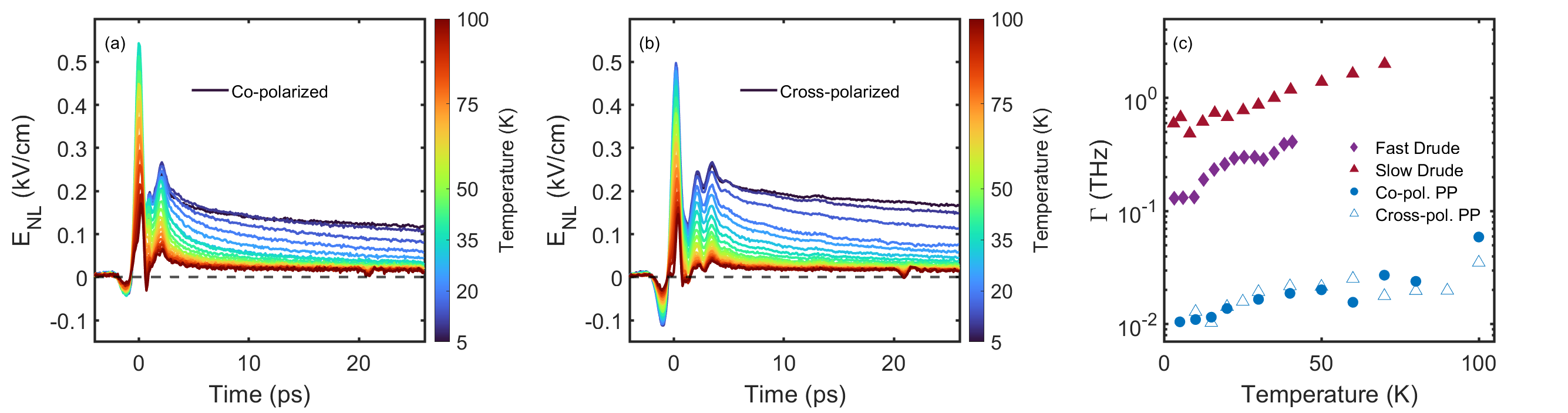}
    \caption{(a) The temperature dependence of the THz pump-probe signal is shown for co-polarized where both pump and probe are at $\theta = 0^{\circ}$ and (b) cross-polarized where pump and probe are at $\theta = \pm45^{\circ} $ THz pulses. (c) Decay rates from fits to the phenomenological bi-exponential model. The pump-probe relaxation rates are plotted alongside the two relevant Drude relaxation rates from linear THz spectroscopy~\cite{wang_separated_2021}.}
    \label{Fig2a_PumpProbeTemperature}
\end{figure*}

In this Letter, we use nonlinear THz 2D coherent spectroscopy (2DCS) to study intrinsic energy relaxation rates in ultraclean Sr$_2$RuO$_4$ thin films. We find the largest $\chi^{(3)}$ response is in the low temperature Fermi-liquid regime, but that the nonlinear response persists up to intermediate temperatures $T \sim 100-200$ K well above the FL coherence temperature. We find that the dominant contribution to the third-order current is due to THz range pump-probe processes that are governed by energy relaxation. The time-dependence of the pump-probe response can be parameterized as a bi-exponential, with a fast decay and a much longer lived decay that is present at low temperature. The temperature dependence of the {\it faster} decay, which may be a measure of energy flow from the excited electrons to the phonon bath, is found to be 25 times smaller than the previously found (already small) Drude momentum relaxation rate at low temperatures. We track the {\it slower} decay out to nearly 400 ps and find that it decays at a rate $\Gamma_{E; slow} \leq 1$ GHz. This feature is most prominent at the lowest temperatures and is consistent with the suppressed decay rates of non-equilibrium phonons and their loss of energy via diffusion.

This study is enabled by new THz pump-probe spectroscopies including 2DCS~\cite{hamm2011concepts,kuehn_two-dimensional_2010,lu_two-dimensional_2018,mahmood_observation_2021} and high field sources. Although there has been extensive work~\cite{groeneveld1995femtosecond,averitt2002ultrafast,russell2023electronic} using optical pump experiments (with photon energies great than 1 eV) on correlated metals, THz pumping offers several simplifications. When applied to optical pumping the TTM model has been criticized on the grounds that its assumption of fast initial thermalization of the electron system is inappropriate~\cite{groeneveld1995femtosecond,kabanov_electron_2008}. For instance, it has been argued that since electron interactions go as $E^2$, thermalization must slow down as it proceeds and must always be taken into account to model energy loss.  We believe that these objections primarily apply to optical pump measurements that use greater than 1 eV photons. In contrast, in our THz experiments (1 THz $ = 4.1$ meV) the electron distribution is not driven far from thermal and we belive differences from equilbrium can be characterized by the deviation of the electron momentum space occupation projected onto a small number of basis functions of the irreducible representations of the material's point group (See Supplemental Material (SM)).  On the time scales of the present experiment we find that the relaxation is dominated by the uniform deviation of the Fermi surface $(A_{1g})$, which is equivalent to energy relaxation.  The development of intense THz sources have enabled recent work studying the nonlinear response of semiconducting, magnetic and superconducting systems~\cite{kuehn_two-dimensional_2010,lu_two-dimensional_2018,mahmood_observation_2021,cheng2023evidence,liu2023probing}; nonetheless, the nonlinear THz response of correlated metals is suprisingly underexplored. In conventional metals with parabolic bands and only elastic scattering the intraband nonlinear susceptibility $\chi^{(3)}$ vanishes.   Older work on the free-electron carrier response in doped semiconductors, shows that well-defined contributions to the $\chi^{(3)}$ response can be attributed to either band non-parabolicity or from energy-dependent scattering~\cite{wolff_theory_1966,rustagi_effect_1970,yuen_differencefrequency_1982}.  In our case either possibility is sufficient to give that the time scale of the non-linearity we observe is largely governed by energy relaxation. While the historical focus in Sr$_2$RuO$_4$ has been on characterizing unconventional superconductivity~\cite{maeno_superconductivity_1994}, we focus on its normal state being an important test case as it is believed to be a quasi-2D Fermi-liquid (FL) for $T_{FL} \leq 25-40$ K (as demonstrated by the presence of a $T^2$ dependent resistivity and $\omega^2$ dependent scattering rate)~\cite{maeno_two-dimensional_1997,stricker_optical_2014,wang_separated_2021}.

We use a standard implementation of THz 2DCS~\cite{kuehn_two-dimensional_2010,lu_two-dimensional_2018,mahmood_observation_2021} with two separate THz pulses with fields $E_A$ and $E_B$ and control of both the gate time $t$ and the relative delay between pulses $\tau$.  
In Fig.~\ref{Fig1_2pulse}(a), we set the relative delay $\tau = 0$ between the two THz pulses and present the measured nonlinear response as the difference in using both pulses together and using each separately $E_{NL}(t) = E_{AB}(t) - E_A(t) - E_B(t)$.  We measure as a function of temperature for a 18.5 nm thick Sr$_2$RuO$_4$ thin film grown by molecular beam epitaxy on a $(110)$ NdGaO$_3$ substrate~\cite{nair_synthesis_2018}.  The measured electric field is proportional to the derivative of the induced time-dependent polarization.  The nonlinear signal increases as the temperature decreases, reaching a maximum at our lowest measured temperature of 5 K.  Notably, the nonlinear response persists to temperatures much beyond the FL regime. The corresponding Fourier transform of the nonlinear response is shown Fig.~\ref{Fig1_2pulse}(b), where we find that the frequency content appears to follow the input intense THz pulses closely (in gray). This indicates a relatively featureless $\chi^{(3)}$. For a broadband pulse, the appropriate normalization scheme for $\chi^{(3)}(\omega)$ is complicated as one needs to consider both the amplitude and phase at each $\omega$ within the applied THz pulse. We show in Fig.~\ref{Fig1_2pulse}(c) the simplest ratio $E_{NL}(\omega)/E_{B}(\omega)$ to normalize for the temperature dependence of the linear transmission. 

Since nonlinear responses can arise from multiple excitation pathways, we use THz 2DCS to identify the origins of observed nonlinearity as shown in Fig.~\ref{Fig1_2pulse}(d,e) at 10 K in a cross-polarized configuration. We find the only appreciable contribution to the nonlinear signal corresponds to the position in the 2DCS spectra (Fig.~\ref{Fig1_2pulse}(e)) of the ``pump-probe" response, which appears as a streak along the $\nu_\tau = 0$ axis (See Supplemental Materials (SM) of~\cite{mahmood_observation_2021} for detailed discussion).  As discussed in the SM, widths of spectral features in the $\hat{\nu_\tau}$ direction are related to the rate that energy leaves the electrons. 

Having identified the principal non-linear contribution to the data as being the ``pump-probe" response, further analysis about decay rates can in principle be done in either the time- or frequency-domain.  Nevertheless, as further discussed in the SM, frequency-domain analysis requires additional consideration and must be ``phase untwisted" before peak widths are analyzed as shown in Fig.~\ref{Fig1_2pulse}(f). Although untwisted data are in excellent agreement with our time-domain fits shown below we prefer time-domain fits to fitting untwisted full 2DCS spectra due to the long amount of time it takes to scan the two axes to take a full 2DCS spectra and the fact that ultimately we have found the time-domain data at fixed $t$ to give a clearer perspective on subtle features relevant here due to its inherently higher single to noise.  Therefore, we consider the temperature-dependence of energy relaxation in Sr$_2$RuO$_4$ using a 1D variant of 2DCS, THz ``pump-probe" (PP) spectroscopy, where we set the delay time $t$ and scan only $\tau$. We use a strong-pump, strong-probe regime in order to directly link our time-resolved experiments to the features in the 2DCS response. In Fig.~\ref{Fig2a_PumpProbeTemperature}(a,b) we show the temperature dependence of the pump-probe response studied with fixed gate time $t$ and varying the relative delay $\tau$ for both co-polarized (pump and probe both at $\theta = 0^{\circ}$) and cross-polarized (pump and probe at $\theta = \pm 45^{\circ}$) geometries.  The magnitude of the response decreases as temperature increases, much as was seen in Fig.~\ref{Fig1_2pulse}. The initial fast oscillations for $\tau < 5$ ps are an artifact arising from PP processes from when pulse $B$ precedes (or is coincident with) pulse $A$ e.g. the roles of the pulses for pump and probe are reversed.  Then a $\tau$ scan at fixed $t$ can give oscillations from scanning across the A pulse.

\begin{figure*}[!ht]
    \centering
    \includegraphics[width = 1.0 \textwidth]{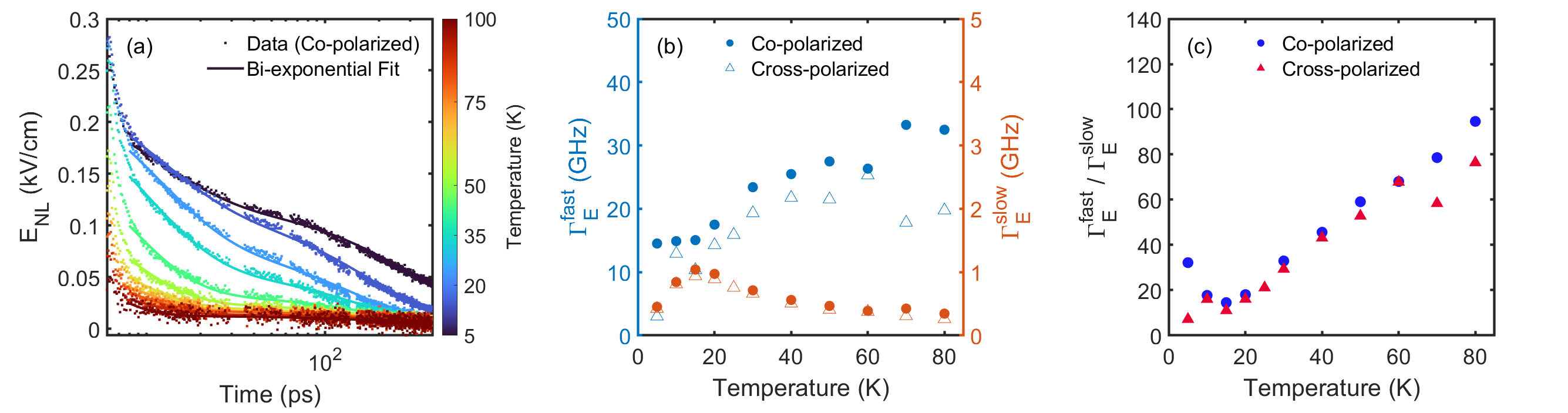}
    \caption{(a) The pump-probe data over an extended scan range up to 380 ps is shown along with a bi-exponential fit for the co-polarized geometry. The cross-polarized data follows similar behavior. There are small time gaps where data were cut to remove reflections of the THz pulse within the detection crystal or within the substrate (the long-term effect of these signals is negligible). (b) Energy relaxation rates for both polarizations. (c)  The ratio of the two relaxation rates to be compared to those commonly extracted in thermoelectric transport measurements showing phonon drag. }
    \label{Fig6_longscans}
\end{figure*}

In order to study the relevant timescales of energy relaxation, we parameterize the data for $t > 5$ ps using a bi-exponential model $    E_{NL}(t)  = A_{fast} \; \mathrm{exp} [-2\pi \,\Gamma_{E,fast} t]  + A_{slow} \; \mathrm{exp} [-2\pi  \,\Gamma_{E,slow} t]$, where  $\Gamma_{E,fast}$ and $\Gamma_{E,slow}$ are the energy relaxation rates for the fast and slow decay processes. We fit the experimental data to this bi-exponential model and plots the rates as a function of temperature in Fig.~\ref{Fig2a_PumpProbeTemperature}(c). Note the $2\pi$ in the definitions above that we have included so as to make comparisons to rates that come via widths of frequency domain features (e.g. Drude conductivity) straightforward.  Fitting exponentials to the time domain data gives the same results as fitting vertical Lorentzian cuts (as in Fig. 1(f)) through untwisted versions of data in Fig. 1(e) as long as the frequency dependence to the widths in the $\nu_t$ direction are weak.   This appears to be the case presently.  We plot in Fig.~\ref{Fig2a_PumpProbeTemperature}(c) the temperature dependence of the faster decay for both polarization geometries.  As discussed in the SM, the co-polarized geometry is sensitive to the decay of both $A_{1g}$ and $B_{1g}$ perturbations to the Fermi surface and the cross-polarized geometry is sensitive to decays of both $A_{1g}$ and $B_{2g}$ decays.   The fact that similar decay rates are seen in both geometries likely shows that we are largely sensitive to only $A_{1g}$ decays, with the $B_{1g}$ and $B_{2g}$ contributions probably having decayed away faster than can be resolved with $\approx$ 1 ps pulses.  $A_{1g}$ decay is equivalent to energy relaxation as it usually will correspond to momentum decay with a contribution in the direction perpendicular to the FS and hence in the direction of the energy gradient.   One expects that this decay then may be described by the TTM.  The decay rate increases with temperature in accord with the TTM's prediction for temperatures well below the Debye temperature~\cite{allen_theory_1987,groeneveld1995femtosecond}.  Within the interpretation that energy is lost to the lattice, one can infer a dimensionless electron phonon coupling constant $\lambda \approx 0.18$ assuming Allen's expression and literature values for the Debye temperature~\cite{allen_theory_1987,nishizaki1998evidence}.   This is a value consistent with what has been found earlier via photoemission~\cite{ingle2005quantitative}.  

One can also compare these rates to other decay rates. The width of the optical conductivity's ``Drude" peak is the measure of the current decay.  Net current is proportional to total momentum for simple band structures.   Current decay is determined by the scattering probability of a quasi-particle state weighted by the fractional change of current from a scattering event~\cite{lundstrom2002fundamentals}. We have previously shown that a number of ruthenates (Sr$_2$RuO$_4$ included) show a two-Lorentzian optical conductivity at low temperature~\cite{wang_separated_2021} that both scale as $T^2$ at low T.  We also plot in Fig.~\ref{Fig2a_PumpProbeTemperature}(c) the $T$ dependence of these rates ($\Gamma_{\sigma,\; fast}$ and  $\Gamma_{\sigma,\; slow}$). One can see that the ``fast" energy relaxation rate is 15-60$\times$ smaller than the optical conductivity scattering rates.

How can we make sense of this apparent mismatch in the scales of the different relaxation processes?   As discussed above, in the conventional view energy relaxation from the electron system arises from a subset of scatterings that loose energy to the lattice. The mismatch between $\Gamma_E$, and  $\Gamma_{\sigma,\; fast}$ and  $\Gamma_{\sigma,\; slow}$ indicates that the scattering processes that lose energy are overwhelmingly not the same as that which leads to current decay in this material.  This is consistent with the current decay largely coming from umklapp and inter-band e-e scattering.

The slow decay rate ($\Gamma_{E; slow} \leq 1$ GHz) found in our PP studies requires further investigation. In Fig.~\ref{Fig6_longscans}(a,b), we show the dynamics of the PP response out to 380 ps for both co-polarized and cross-polarized geometries.  In Fig.~\ref{Fig6_longscans}(c), we show the rates as a function of temperature from fits to the bi-exponential model.  Among other aspects there is a peak in the decay rate $\Gamma_{slow}$ around 20 K - how should think to understand this very long time scale?  Similar long time dynamics have been observed in superconductors due the formation of a phonon bottleneck and modeled using Rothwarf-Taylor dynamics~\cite{rothwarf_measurement_1967}.  The particularities of that effect do not apply in the present case, but in general long decay can arise through a non-equilibrium phonon population, whereby the re-excitation of the electron system by non-equilibrium phonons inhibits electron relaxation~\cite{bailyn_transport_1958,m_bailyn_phonon-drag_1967,r_n_gurzhi_electric_1972, cai_nonequilibrium_1987,yu_g_gurevich_electron-phonon_1989, wyatt_suprathermal_2000,adamenko_suprathermal_2003}.  We model this is in the SM.  One of the typical signatures of non-equilibrium phonons is a peak in the thermopower at low $T$ in very clean conductors. It can be present even if there is no noticeable change in the resistivity. Thermopower measurements in Sr$_2$RuO$_4$ have shown an anomalous upturn in $dS/dT$ at 20-25 K, which did not correspond to any change in other transport coefficients~\cite{yoshino_thermopower_1996,xu_band-dependent_2008} and may point to the role of non-equilibrium effects at low $T$ in Sr$_2$RuO$_4$.  It is likely to be complicated to model the thermopower in Sr$_2$RuO$_4$ due to the presence of three bands~\cite{cuoco_thermopower_2003}, but we can note the upturn in  $dS/dT$ coincides with the temperature where the peak in the long time decay is found.  The relaxation dynamics of the electron-phonon system is ultimately governed by thermal diffusion and in this regard we tentatively assign this long time scale of energy decay out of the electron system as being limited by phonon diffusion.  The effective impedance mismatch between the film and substrate phonons could perhaps being decreasing the efficiency of energy diffusion.   It is also interesting to note that the amplitude of the slow decay seems to scale linearly with pump field (see SM).   This would seem to indicate an interface where inversion symmetry is broken is at play.

We have found that the large nonlinear THz response in Sr$_2$RuO$_4$ is tied to energy relaxation processes. We observe two energy relaxation rates: the energy relaxation rate from the electron to the phononic subsystem and a longer rate that we believe is set by the diffusion of nonequlibrium phonons. Further work on other ultraclean correlated metals would help clarify the role of different electronic scattering mechanisms.  Moreover, as they are measures of viscosity~\cite{alekseev2016}, it would be interesting to extend these studies to even shorter time scales with shorter pulses to see if $B_{1g}$  and $B_{2g}$ decays can be resolved.  Finally, although we have a qualitative understanding of our data, our work also highlights how underdeveloped the theory is and the concomitant possibilities for an improved theoretical understanding of such processes in interacting systems.

The authors thank P. Allen, K. Behnia, L. Benfatto, G. Grissonnanche, S. Hartnoll, V. V. Kabanov, K. Katsumi, G. Kotliar, D. Maslov, L. Taillefer, and P. Volkov for helpful discussions.   The project at JHU was supported by the NSF-DMR 2226666 and the Gordon and Betty Moore Foundation’s EPiQS Initiative through Grant No. GBMF9454. NPA had additional support by the Quantum Materials program at the Canadian Institute for Advanced Research. Research at Cornell was supported by the National Science Foundation [Platform for the Accelerated Realization, Analysis, and Discovery of Interface Materials (PARADIM)] under Cooperative Agreement No. DMR-2039380. This research was also funded in part by the Gordon and Betty Moore Foundation’s EPiQS Initiative through Grant Nos. GBMF3850 and GBMF9073 and NSF DMR-2104427 and AFOSR FA9550-21-1-0168.  Sample preparation was, in part, facilitated by the Cornell NanoScale Facility, a member of the National Nanotechnology Coordinated Infrastructure (NNCI), which is supported by the National Science Foundation Grant No. NNCI-2025233.

\bibliography{Ruthenates}
 
\clearpage
\pagebreak
\widetext
\begin{center}
\textbf{\large Supplemental Material: Energy relaxation and dynamics in the correlated metal Sr$_2$RuO$_4$ via two-dimensional THz spectroscopy}
\end{center}

\section{Methods} 
The $(001)$ Sr$_2$RuO$_4$ thin films were grown via MBE on (110) NdGaO$_3$ substrates. The films were characterized via XRD and HAADF-STEM to ensure homogeneity. More details on their fabrication can be found elsewhere~\cite{nair_synthesis_2018}. Due to the supression of ruthenium vacancies by growing with an overpressure of ruthenium oxide under adsorption-controlled conditions in MBE the residual resistivity ratio (which we define as $\rho(300 K)/\rho(4 K)$ was $>50$, which is an exceptional value for thin film transition-metal oxide films. As such, these Sr$_2$RuO$_4$ films represent an excellent system to study the optical response of a strongly correlated clean Fermi liquid in the clean limit. 

The two co-linear THz pulses used for nonlinear spectroscopy were generated in two separate LiNbO$_3$ crystals and then recombined.  The electric field strength of each THz pulse individually is $\sim 40$ kV/cm. The THz pulses were detected using electro-optic sampling in a 0.5 mm GaP crystal for the 2DCS and the nonlinear spectroscopy experiments. As the signal in the pump-probe experiments was much smaller, a 0.5 mm or 1.0 mm ZnTe crystal was used. In the co-polarized regime, the electric field was applied  $E \parallel [\overline{1}10]$ on the substrate. The two THz beams could be oriented such that the two beams with co-polarized $\theta = 0^\circ$ parallel to the $[\overline{1}10]$ axis of the detection crystal,  or cross-polarized at $\theta = \pm 45^\circ$ with respect to the $[\overline{1}10]$ axis on the detection crystal. 

A number of different kinds of nonlinear THz experiments were carried out to study the response of carriers in Sr$_2$RuO$_4$. The most general experiment is THz 2D coherent spectroscopy (2DCS) where the nonlinear response is obtained as a function of gate time $t$ and the delay time between pulses $\tau$. Due to the fact that two time axes are scanned, this is a very time-intensive method, so in order to extract the temperature-dependence, we used THz pump-probe experiments that are related to the full 2DCS experiment for fixed gate time $t$ and variable delay time $\tau$. Similarly, the nonlinear response proportional to the overall $\chi^{(3)}_{eff}$ can be mapped out by setting $\tau = 0$ and observing the nonlinear response resulting from the overlap of the two THz pulses. More specific details on this implementation of 2DCS can be found elsewhere~\cite{mahmood_observation_2021}.

\section{Partial Phase Untwisting}

Phase untwisting in the data analysis of frequency domain $\chi^{(3)}$ is necessary to reveal strictly dissipative line shapes as the straightforward application of 2D spectroscopy will yield a mix of dissipative and reactive contributions in the real and imaginary $\chi^{(3)}$~\cite{hamm2011concepts}.  The procedure for phase untwisting in 2DCS is relatively straightforward if features are narrow in the frequency domain and only a single non-linear excitation pathway dominates, which in the present case is the pump-probe response~\cite{kuehn_two-dimensional_2010}.    We follow the methods outlined in  Ref.~\cite{hart_extracting_2022}, by first symmetrizing the time-dependent data $E_{NL}(t,\tau$) over $\tau$. In the raw 2DCS spectra, we cut the data for $\tau < 1.0$ ps to remove contributions from the inverted AB pulse ordering that arises from the finite temporal extent of the pulse.    We then let $\tau \rightarrow -\tau$ and then add the original and inverted data to each other and then take the Fourier transform.   Because the effective time domain data are now (by construction) a symmetric function, it is purely real.   We can then take the magnitude of the resulting quantity which is then (if the decay was purely exponential) a perfect Lorentzian.  Its width should be related to the inverse lifetime of the relevant excitation.

\begin{figure}[t]
    \centering
    \includegraphics[width = 0.5\columnwidth]{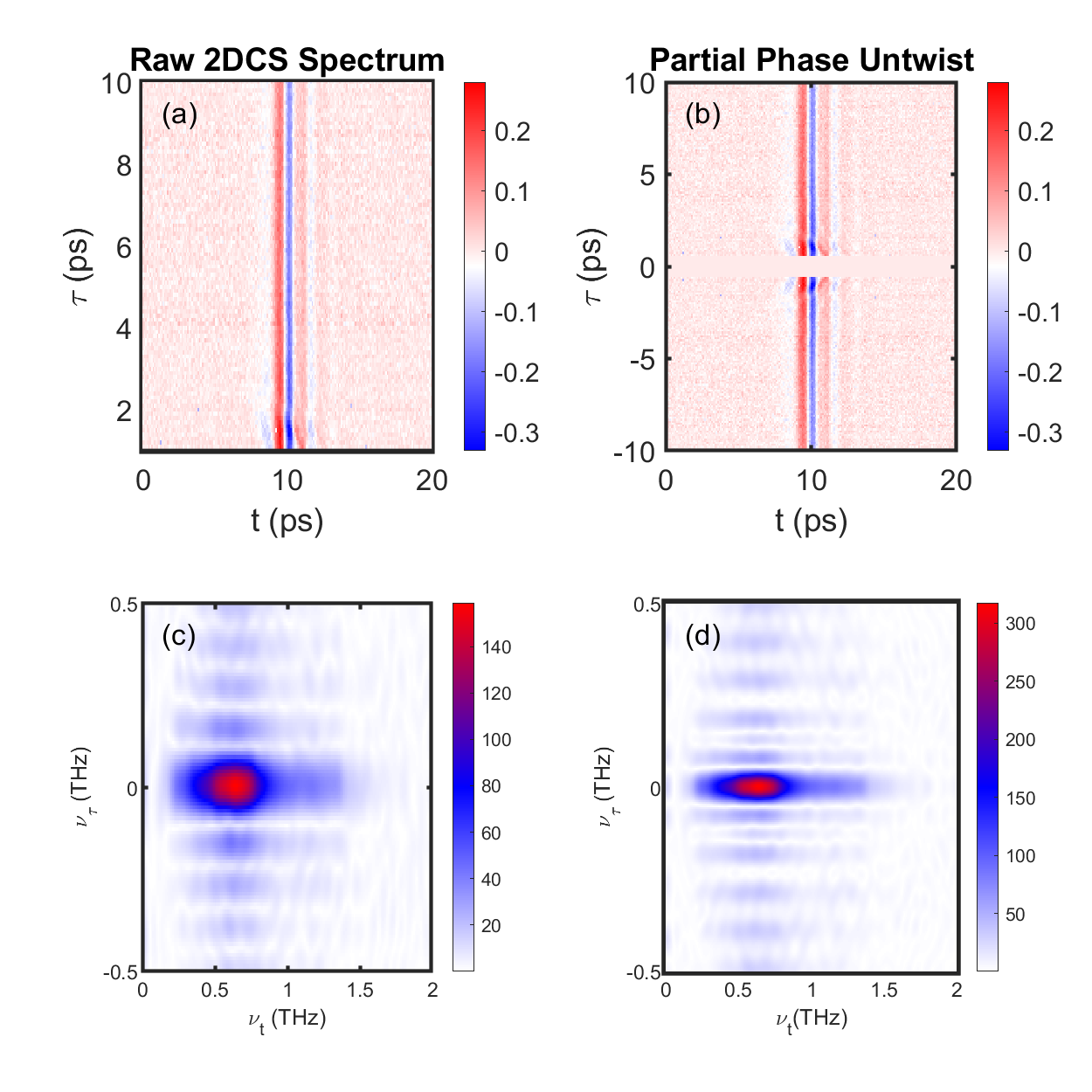}
    \caption{The raw 2DCS time-traces and spectrum (a,c) as compared to the resulting untwisted time-traces and spectra (b,d).  The spectral features are found to be much sharper in the untwisted spectra due to a lack of mixing between parts of the complex response of the spectrume.}
    \label{SM_2DPhaseUntwist}
\end{figure}

In addition to correcting the phase, this procedure also acts like a high-pass filter, which results in some high-frequency artifacts away from the main Lorentzian-like peak. Fortunately, this procedure gives a much narrower peak near $\omega_B = 0$ that is reflective of the intrinsic inverse lifetime.

\section{Nonlinear THz experiments as a measure of energy relaxation}

It is important to work out more carefully what the decay rates that are being probed corresponded to.   A general model of this decay does not yet exist, but we can gain intuition about what is measured in these THz pump-probe experiments by considering the third order non-linearity as an effective linear response of a system in quasi-equilibrium that is slowly decaying after being pumped.   This picture should be valid if the decay rates are longer than the time scales of the pulses.  The nonlinear $\chi^{(3)}$ response is generally defined as   

\begin{equation}
P_i = \chi^{(3)}_{ijkl} E_l  E_k  E_j.
\end{equation}

Consider a situation similar to the present one where $E_l $ and $ E_k $ derive from the same pulse e.g. $E_l =  E_k $ and precede $E_j$ in time.  If -- after a perturbation $E_l^*E_l$ -- the excited system decays over a time scale much longer than the time that the pulse $E_j$ interacts with the sample, then we can consider the resulting sample polarization of the system in the quasi-equilibrium state to be  

\begin{equation}
P_i = \chi^{(1)'}_{ij}  E_j,
\label{Polarization}
\end{equation}
where the effective quasi-equilibrium $ \chi^{(1)'}$ is

\begin{equation}
 \chi^{(1)'}_{ij}= \chi^{(3)}_{ijll} |E_l|^2.
\end{equation}
This object $\chi^{(1)'}_{ij}$ is clearly less symmetric than the original $ \chi^{(3)}_{ijkl}$.  For a $ \chi^{(3)}_{ijkl}$ of a crystal with a $D_{4h}$ symmetry and $E_l$ pointing at an angle $\phi$ with respect to the $\hat{x}$ direction, $\chi^{(1)'}_{ij}$ can be expanded in the basis functions of the irreducible representations of the D$_{4h}$ point group.   The general form for the $A$ before $B$ pulse sequence in the time-domain is

  \begin{align}
\chi^{(1)'}(\tau,t) \propto  e^{ - 2 \pi \Gamma_{E_{1u} } t  }  H(\tau) \; \; &  \times \nonumber \\ \Bigg ( \delta_{A_{1g} } e^{ - 2 \pi \Gamma_{A_{1g} } \tau  } & \; \left[\begin{array}{cc}1 & 0 \\ 0 & 1 \end{array}\right]  +  \delta_{ B_{1g} } e^{ - 2 \pi \Gamma_{B_{1g} } \tau  }   \;  \mathrm{cos} 2 \phi   \left[\begin{array}{cc}1 & 0 \\ 0 & -1 \end{array}\right] +   \delta_{ B_{2g} } e^{ - 2 \pi \Gamma_{B_{2g} } \tau  }   \;   \mathrm{sin} 2 \phi  \left[\begin{array}{cc} 0 & 1 \\ 1 & 0 \end{array}\right]   \Bigg ).
\label{Response}
\end{align}
 Here $H(\tau)$ is a step function.  In the quasi-equilibrium picture suggested above, the $\delta$'s in Eq. \ref{Response} represent deviations of the electron momentum space occupations from equilibrium projected onto the basis functions.     The different contributions can be understood as follows.  $ \delta_{A_{1g} } $ represents uniform deviations of the electrons to states of greater momentum.   For simple bandstructures, finite $ \delta_{A_{1g} } $   gives a state of greater energy and hence its time dependence gives the rate of energy decay.  $ \delta_{E_{1u} } $ is a horizontal shift of the momentum and represents a configuration that carries current.  $ \delta_{B_{1g} } $  is a ``$d_{x^2-y^2}$"-like deviation with greater occupation of states at $\vec{k}_F$ for the Fermi surface crossing in the $\hat{x}$ direction and lesser occupation for states in the $\hat{y}$ direction. $ \delta_{B_{2g} } $ is a similar deviation, but rotated by 45$^\circ$ e.g. $d_{xy}$-like.  At $\tau = 0^+ $, the deviations of the momentum space occupations reflects projection of the $|E_l|^2 \mathrm{cos} 2\phi $ perturbation onto the basis functions.   Therefore in the impulsive limit, when $E_l$ have time dependencies that approach delta functions, then  $\delta_{A_{1g} }  = \delta_{B_{1g} }  =  \delta_{B_{2g} } $, because to lowest order a perturbing electric field in the $\hat{\phi}$ direction cannot give an $E^2$ perturbation at $\tau = 0^+$ to the near-$E_F$ occupations for states whose $\vec{k}_F$ is perpendicular to $\hat{\phi}$ and so the different contributions must have equal amplitude at $\tau = 0^+$.  Note again the factor of $2 \pi$ in the exponentials that we include so that the rates can be straightforwardly compared with the widths of spectral features that are arrived at by Fourier transform.   The nonlinear emitted electric field is proportional to the time derivative of Eq. \ref{Polarization}.
 
 In the experiment detailed in the main text we see essentially no anisotropy in the polarization dependence.  As there is no conservation law (e.g. energy or momentum) for $B_{1g}$ and $B_{2g}$ decays (as there are for $A_{1g}$ and $E_{1u}$) it reasonable to expect that  $ \Gamma_{A_{1g} } \ll \Gamma_{B_{1g} }, \Gamma_{B_{2g} }  $ and therefore at the measurement times $\tau> 3$ ps resolvable in the present experiment the $B_{1g}$ and $B_{2g}$ deviations have already decayed away leaving only the energy decay ($A_{1g}$) measurable.  In order to resolve  $B_{1g}$ and $B_{2g}$  decays, it is likely that one needs to use shorter pulses.

\section{Toy Model of Energy Transfer}
We can establish a toy model to understand qualitative aspects of our THz pump-probe experiment and energy decay. We use the formalism of Ref.~\cite{yu_g_gurevich_electron-phonon_1989} and consider the following kinetic equations. At times after the THz pulse is no longer interacting with the system (e.g. neglecting the forcing function of the external electric field), the electron and acoustic phonon systems can be governed by the following differential equations,
\begin{equation}
    \begin{split}
        \frac{\partial f}{\partial t} + \mathbf{v}\frac{\partial f}{\partial \mathbf{r}} & = S_{ep} + S_{ed} + S_{ee} \\
        \frac{\partial N}{\partial t} + \mathbf{v}_q \nabla N & = S_{pe} + S_{pp} + S_{pd} 
    \end{split}
\end{equation}
where $f,N$ are the distribution functions of the electrons and acoustic phonons, $\mathbf{v},\mathbf{v}_q$ are electron and phonon velocities respectively, and $S_{ik}$ are the respective collision intervals between the scattered object ($e,p$ denote electron and phonon respectively) and the scatterer ($e,p,d$ denote electron, phonon or disorder). We assume that the change to the electron and phonon distributions after interaction with the THz pulse is a small perturbation.

The form of the electron distribution depends on the relative hierarchy of scattering rates between the momentum relaxation rate $\Gamma_\sigma$, the electron-electron scattering rate $\Gamma_{ee}$ (which governs the rate of electron thermalization) and the electron energy relaxation rate $\Gamma_E$ that governs the rate of energy leaving the electron subsystem. We make the assumption that due to the low frequency of the THz pump the electron distribution is never driven that far from a thermal one.   As discussed above, at most the deviation from equilibrium can be characterized by the projection of small changes in occupation near ${\bf k}_F$ onto the basis functions of the irreducible representations of the crystal point group.  The fact that we see no angular anisotropy on the time scales probed in this experiment, indicates that we can characterize the data presented here in terms of fully symmetric $A_{1g}$-like deviation of the quasiparticle distribution from equilibrium.   For simple Fermi surfaces, such a deviation is only relaxed by movement of the quasiparticles~\cite{rustagi_effect_1970} in a direction that projects on ${\bf k}_F$ and so $A_{1g}$ like relaxation is equivalent to energy relaxation.

To understand the observed time-dependence of the pump-probe response, we model our system with a modified two-temperature model (TTM) only considering low energy acoustic phonons~\cite{perfetti_ultrafast_2007} instead of solving the full Boltzmann equations presented above. We take the limit $\Gamma_{ep} \ll \Gamma_{pp}$ and neglect the reverse $\Gamma_{pe}$ phonon absorption process, which should be appropriate in the limit of low temperatures.  We treat the THz pump in the perturbative limit and focus on the long-time behavior, which avoids some of the limitations of the TTM at low temperatures~\cite{kabanov_electron_2008}. We also assume that $T_e \gtrsim T_p$ in the regime of a perturbative excitation by the THz pulse, where $T_p$ describes the temperature of a subset of coupled phonons that can be at a different temperature than the rest of the system.  The assumption is that on long time scales the heat in the phonon system will diffuse to a heat sink that is the substrate and the cryostat ultimately.  The time dependence given by our toy model is governed by the coupled equations:
\begin{equation}
    \begin{split}
        \frac{\partial T_e}{\partial t} & = -A\, \Gamma_{ep}(T_e-T_p) \\
        \frac{\partial T_p}{\partial t} & = \frac{C_e}{C_p}A\, \Gamma_{ep}(T_e-T_p) - B \, \Gamma_{ps} (T_p - T_{sink}), \\
    \end{split}
\end{equation}
where $\Gamma_{ep}$ is the energy relaxation rate from electron-phonon scattering and $\Gamma_{ps}$ is the rate of energy loss via heat diffusion to a heat sink, and $C_e, C_p$ are the specific heat coefficients of the electrons and the coupled phonon system. 

\begin{figure}[t]
    \centering
    \includegraphics[width = \textwidth]{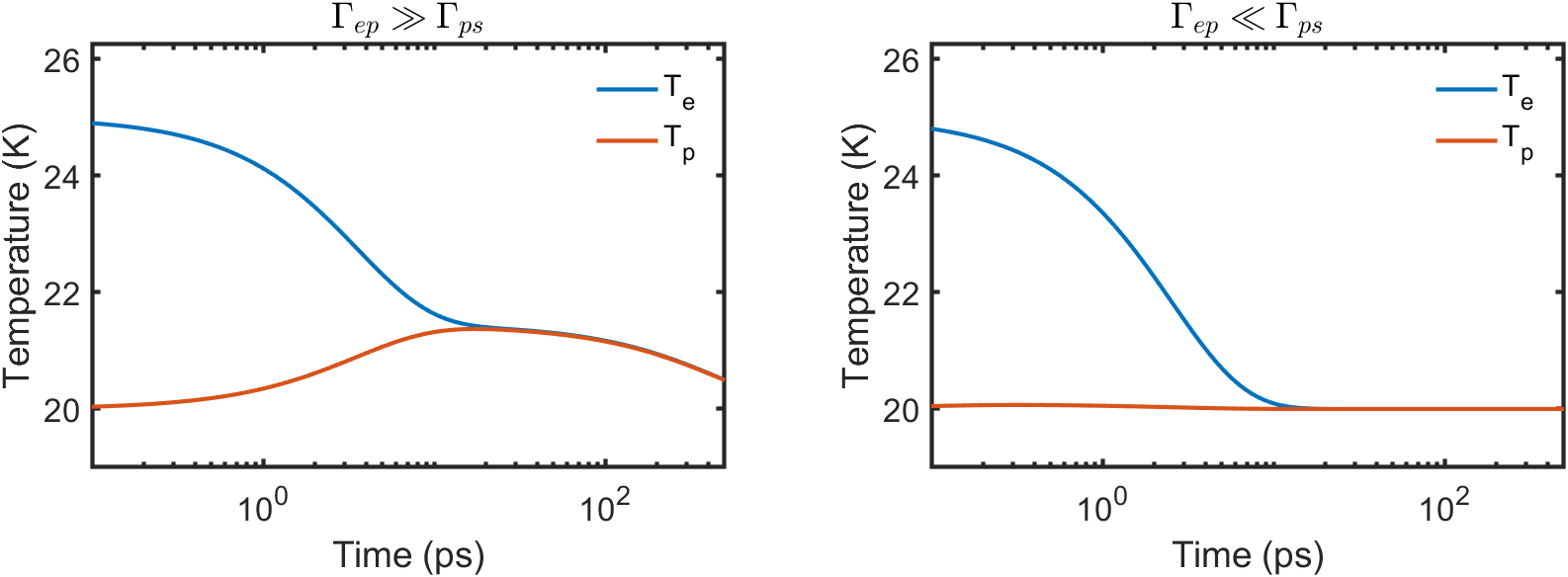}
    \caption{Simulations of two different limits for the electron-phonon and phonon-phonon energy relaxation rates. }
    \label{SM_ToyModel}
\end{figure}

As the temperature increases, the number of energy diffusing phonons increases.  They are able carry off excess energy in the phonon distribution, leading to the disappearance of the long lived non-equilibrium regime.  At sufficiently high temperatures, we would expect the non-equilibrium phonon regime to disappear.  Then the pump-probe response is only governed by energy relaxation dynamics by the electron-phonon scattering rate.

In Fig.~\ref{SM_ToyModel}, we show the numerical solution to the above differential equations for two extreme limits of the decay rates: $\Gamma_{ep} \gg \Gamma_{ps}$ and $\Gamma_{ep} \ll \Gamma_{ps}$. We keep $\Gamma_{ep} = 0.1$ THz fixed, and set $\Gamma_{ps} = 1$ GHz and $\Gamma_{ps} = 1$ THz respectively. We find that in the first limit of  $\Gamma_{ep} \gg \Gamma_{ps}$, the time-dependence of the electron temperature captures many of the qualitative features observed in our experimental data at low temperature with the two separate timescales visible. In the opposite limit of $\Gamma_{ep} \gg \Gamma_{ps}$, there is no accumulated energy transferred to the coupled phonon system (no bottleneck) and the rate of energy transfer to the sink is determined by $\Gamma_{ep}$.

\section{Electric field dependence of the Pump Probe Response } 

\begin{figure*}[!ht]
\centering
\includegraphics[width = \textwidth]{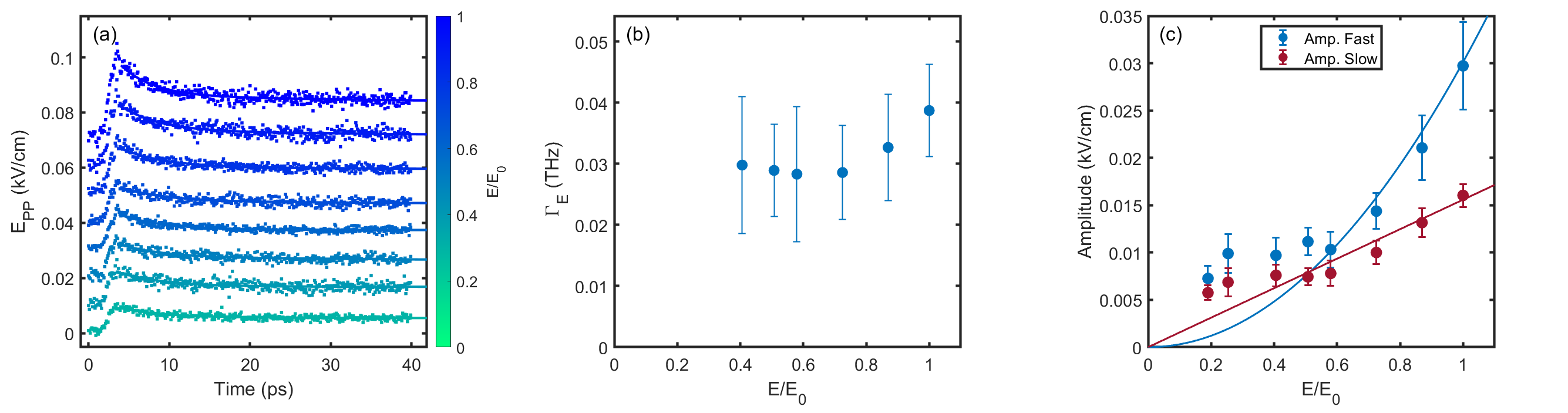}

\caption{The electric field dependence of the pump-probe response is shown at 10 K, with the applied probe field set to $E_{probe} = 5 $ kV/cm and the maximum $E_{pump} = $  50 kV/cm. The two pulses are cross-polarized. The absolute response is shown in (a) with arbitrary offsets added for clarity, with fits to the bi-exponential model from Fig.~\ref{Fig2a_PumpProbeTemperature}(c) also shown. We find that the fast decay channel shown in (b) does not change significantly with a large variation in applied field. We find that the amplitude of the fast decay channel appears to be proportional to $A_{fast} \propto E^2$, whereas the amplitude of the slow decay channel varies only as $A_{slow} \propto E$.} 

\label{Fig3_Fluence}
\end{figure*}

To confirm that the the fields used in our experiment are in the perturbative limit, we studied the field dependence at 10 K in a cross-polarized geometry with the probe fixed at 5 kV/cm and the pump varied between 5 - 29 kV/cm as shown in Fig.~\ref{Fig3_Fluence}(a). Fitting the pump-probe response to the same bi-exponential model, we observe that the fast relaxation rate shown in Fig.~\ref{Fig3_Fluence}(b) appears nearly independent of the applied field. This indicates that the energy relaxation of the quasiparticles probed is in the perturbative limit. In Fig.~\ref{Fig3_Fluence}(c) the field-dependence of the amplitude of the two pump-probe contributions is shown. We find that the amplitude of the fast decay is proportional to pump field square, but the amplitude of the slow decay is linearly proportional to the pump field. Since the fast component governs the decay of excited quasiparticles, we expect it to scale with intensity $\Delta n \propto E^2$. 

\end{document}